\documentclass[10pt,twocolumn,twoside]{IEEEtran}
\usepackage{ifpdf}
\usepackage{url}
\ifpdf
\else
\fi
\usepackage{pifont}
\usepackage{cite}
\usepackage{balance}
\usepackage{bm,comment,color}

\ifCLASSINFOpdf
  \usepackage[pdftex]{graphicx}
  \graphicspath{{../pdf/}{../jpeg/}}
  \DeclareGraphicsExtensions{.pdf,.jpeg,.png}
\else
  \usepackage[dvips]{graphicx}
  \graphicspath{{../eps/}}
  \DeclareGraphicsExtensions{.eps}
\fi
\usepackage{amsmath}
\usepackage{algpseudocode}
\algtext*{EndWhile}
\algtext*{EndIf}
\algtext*{EndFor}
\usepackage{algorithm}
\usepackage{array}
\usepackage{makecell}
\usepackage{amsfonts} 
\usepackage{amssymb}
\usepackage{esint} 
\usepackage{units}
\usepackage{multirow}
\usepackage{comment}

\usepackage{amsthm}

\usepackage{color}

\usepackage{standalone}

\usepackage{pgfplots}
\usepackage{tikz}
\usetikzlibrary{decorations.pathreplacing}
\usetikzlibrary{calc}
\makeatletter
\newcommand{\gettikzxy}[3]{%
  \tikz@scan@one@point\pgfutil@firstofone#1\relax
  \edef#2{\the\pgf@x}%
  \edef#3{\the\pgf@y}%
}
\usetikzlibrary{spy,backgrounds}
\pgfplotsset{compat=newest}
\usetikzlibrary{plotmarks}
\usetikzlibrary{arrows.meta}
\usepgfplotslibrary{patchplots}
\usepackage{grffile}
\newlength\fheight 
\newlength\fwidth 
\usepgfplotslibrary{fillbetween}

\usepackage{acronym}

\acrodef{6g}[6G]{the sixth generation}
\acrodef{ae}[AE]{autoencoder}
\acrodef{lqr}[LQR]{linear quadratic regulator}
\acrodef{aoa}[AOA]{angle-of-arrival}
\acrodef{bs}[BS]{base station}
\acrodef{bse}[BSE]{beam squint effect}
\acrodef{crb}[CRB]{Cram\'er-Rao bound}
\acrodef{dt}[DT]{digital twin}
\acrodef{elaa}[ELAA]{extremely large antenna array}
\acrodef{ff}[FF]{far-field}
\acrodef{gru}[GRU]{gated recurrent unit}
\acrodef{isac}[ISAC]{integrated sensing and communication}
\acrodef{las}[L\&S]{localization and sensing}
\acrodef{los}[LOS]{line-of-sight}
\acrodef{nf}[NF]{near-field}
\acrodef{nlos}[NLOS]{non-line-of-sight}
\acrodef{ofdm}[OFDM]{orthogonal frequency division multiplexing}
\acrodef{ris}[RIS]{reconfigurable intelligent surface}
\acrodef{scc}[SCC]{sensing–communication–control}
\acrodef{sns}[SNS]{spatial non-stationarity}
\acrodef{swm}[SWM]{spherical wave model}
\acrodef{siso}[SISO]{single-input single-output}
\acrodef{ue}[UE]{user equipment}
\acrodef{dmimo}[D-MIMO]{distributed MIMO}
\acrodef{ppo}[PPO]{proximal policy optimization}

\usepackage{tabu,longtable}

\ifCLASSOPTIONcompsoc
 \usepackage[caption=false,font=normalsize,labelfont=sf,textfont=sf]{subfig}
\else
 \usepackage[caption=false,font=footnotesize]{subfig}
\fi
\usepackage{stfloats}
\usepackage[hidelinks]{hyperref}
\usepackage{xcolor}
\long\def\comment#1{}

\newfont{\bbb}{msbm10 scaled 700}

\newfont{\bb}{msbm10 scaled 1100}

\definecolor{darkred}{RGB}{180,0,0} 
\usepackage{graphicx}      

\usepackage{booktabs}

\usepackage{amsthm}

\begin{document} 
\bstctlcite{IEEEexample:BSTcontrol}

\title{Foundation Models for Wireless Localization: Pretraining, Adaptation, and Utilization}

\author{
Guangjin Pan, \IEEEmembership{Member, IEEE}, Jiajia Guo, \IEEEmembership{Member, IEEE}, Zheng Xing, \IEEEmembership{Member, IEEE}, Hui Chen, \IEEEmembership{Member, IEEE},\\  
Chao-Kai Wen, \IEEEmembership{Fellow, IEEE}, Shi Jin, \IEEEmembership{Fellow, IEEE}, Henk Wymeersch, \IEEEmembership{Fellow, IEEE}
\thanks{Guangjin Pan, and Henk Wymeersch are with Department of Electrical Engineering, Chalmers University of Technology, 41296 Gothenburg, Sweden (e-mail: {guangjin.pan; henkw}@chalmers.se).}
\thanks{Jiajia Guo is with the Department of Electronic and Computer
Engineering, The Hong Kong University of Science and Technology, Hong
Kong (Email: jiajiaguo@seu.edu.cn).}
\thanks{Zheng Xing is with College of Computer Science and Software Engineering, Shenzhen University, Shenzhen, 518060, Guangdong, China (e-mail: zhengx@szu.edu.cn).}
\thanks{H. Chen is with the Department of Electrical Engineering, Uppsala University, SE-75105 Uppsala, Sweden (e-mail: hui.chen@angstrom.uu.se). (Corresponding author: Hui Chen).}
\thanks{Chao-Kai Wen is with the Institute of Communications Engineering, National Sun Yat-sen University, Kaohsiung 80424, Taiwan (e-mail:
chaokai.wen@mail.nsysu.edu.tw).}
\thanks{Shi Jin is with the National Mobile Communications Research Laboratory, Southeast University, Nanjing 210096, China (e-mail: jinshi@seu.edu.cn).}
}

\maketitle

\begin{abstract}
Accurate wireless localization is a key enabler for 6G networks, yet remains challenging under diverse and rapidly changing propagation conditions. Model-based methods degrade when multipath channels are non-resolvable and model mismatches occur, while supervised deep learning demands large labeled datasets and generalizes poorly to new deployments. Inspired by foundation models (FMs) in language and vision, this article presents a unified framework for FM-based wireless localization that learns transferable channel representations from large-scale unlabeled channel state information and adapts to new environments with minimal or even no supervision. We review the fundamentals of FMs, compare the FM paradigm with existing localization approaches, and introduce a three-stage framework spanning large-scale pretraining, localization-oriented fine-tuning, and context-augmented inference, together with the location-aware applications it enables. Ray-tracing-based case studies show improved positioning accuracy and cross-environment generalization. Finally, we present an outlook on key research directions toward AI-native networks for wireless localization.
\end{abstract}

\begin{IEEEkeywords}
Wireless localization, foundation models, AI-native networks, 6G.
\end{IEEEkeywords}

\IEEEpeerreviewmaketitle
\acresetall 
\section{Introduction}

Accurate wireless localization is a critical enabler for sixth-generation (6G) networks, where it plays two complementary roles. On the one hand, it underpins a wide range of location-based services and emerging vertical applications, including autonomous driving, smart factories, the low-altitude economy, and large-scale internet of things (IoT) \cite{chen2022tutorial}. On the other hand, accurate position information feeds back into the wireless network itself, enabling location-aware optimization such as predictive beamforming, proactive handover, and mobility-aware resource allocation that improve communication efficiency and reliability \cite{pan2025ai}.

Achieving such accuracy under the heterogeneous and rapidly evolving conditions of real wireless networks remains fundamentally challenging. Classical model-based methods estimate geometric parameters such as time-of-arrival (ToA) and angle-of-arrival (AoA) using parametric channel models, but their performance degrades due to a large number of unknowns introduced by non-line-of-sight (NLoS) environments. Data-driven supervised approaches bypass explicit channel modeling by learning direct mappings from channel state information (CSI) to positions. However, they demand large labeled datasets and tend to overfit to specific environments, failing when deployed in new settings \cite{pan2025ai}.

Meanwhile, the rise of foundation models (FMs) in artificial intelligence, exemplified by large language models (LLMs) and vision transformers, has fundamentally reshaped how machine learning systems are built~\cite{fei2022towards}. Instead of training a separate model for each task, FMs are first pretrained on massive unlabeled data using self-supervised learning (SSL) to acquire general-purpose representations, and then adapted to downstream tasks with minimal supervision. Motivated by these strengths, FMs have recently begun to permeate wireless communications \cite{guo2026large}. Early efforts have explored large-scale channel pretraining and self-supervised representation learning for physical-layer tasks such as channel estimation, CSI feedback, beam management, and channel prediction, showing that a single pretrained backbone can benefit multiple tasks~\cite{jiang2025towards}. In particular, for wireless localization, the large wireless localization model~\cite{pan2025large} leverages a pretraining-based FM to improve both positioning accuracy and cross-environment generalization. Similarly, the work in~\cite{liu2026self} explores a masked-autoencoder-based FM to enable self-supervised CSI representation learning for localization. More recently, the study in~\cite{si2025cross} introduces a foundation localization model focusing on indoor, multi-modal localization. Although it discusses the advantages of employing FMs for wireless localization, it does not detail how such an FM is pretrained and adapted, nor the utilization techniques through which external context can be injected at inference time. Building on and complementing these efforts, this article takes a channel-centric perspective and organizes FM-based wireless localization along the pretraining–adaptation–utilization lifecycle~\cite{zhao2026survey}, with emphasis on how self-supervised pretraining extracts transferable channel features and how these features, together with injected context, enable localization under variable wireless deployments.

Building on this insight, we first review the foundational principles of FMs and situate the FM paradigm against existing localization approaches, highlighting its advantages in cross-environment generalization, label efficiency, and task reusability. We then propose an FM-based localization framework that instantiates the three-stage FM lifecycle for positioning. It progresses from large-scale channel pretraining on diverse unlabeled CSI, to localization-oriented fine-tuning with limited labeled data, and to context-augmented inference that injects site-specific knowledge, local databases, and multi-modal observations. The resulting estimates then feed location-aware applications spanning wireless network optimization and vertical services.
To validate the framework, we conduct case studies showing that diverse pretraining substantially improves localization accuracy and that retrieval-based context augmentation effectively closes the accuracy gap in unseen environments. Finally, we discuss how this framework fits into the broader vision of AI-native wireless networks and outline an outlook on the key research directions for the future.

\section{FM as a New Paradigm for Wireless Localization}

\subsection{Fundamentals of Foundation Models}
\label{subsec:fundamentals}

\begin{figure}[t]
    \centering
    \includegraphics[scale=0.4]{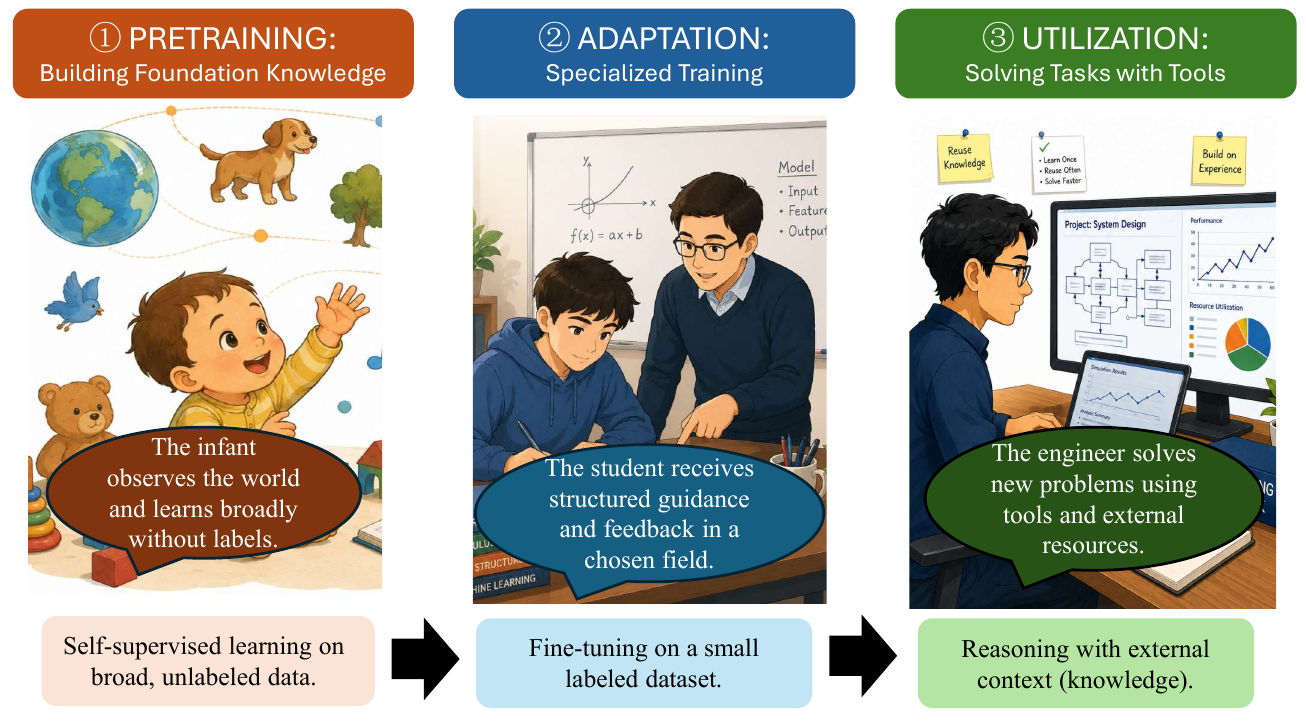}
    \caption{The three-stage lifecycle of FMs (pretraining, adaptation, and utilization), illustrated as a learning journey from a curious child to a tool-augmented expert.}
    \label{fig:fm_augment}
    \vspace{-4mm}
\end{figure}

FMs are pretrained neural networks that leverage SSL on broad, diverse data to obtain general-purpose representations adaptable to a wide range of downstream tasks~\cite{zhao2026survey}. The lifecycle of an FM can be organized into three stages: \emph{pretraining} the encoder on massive unlabeled data to learn general-purpose representations, \emph{adaptation} to specialize the model for a specific task through weight updates, and \emph{utilization} at inference time through context injection without weight updates. As summarized in Fig.~\ref{fig:fm_augment}, this lifecycle captures the progression of an FM from broad knowledge acquisition, through task-specific specialization, to inference-time problem solving with external context or knowledge. {This progression mirrors a human learning journey: from an infant who acquires broad knowledge of the world through unlabeled observation, to a student who receives formal education with structured guidance and feedback in a chosen field, and finally to a practicing engineer who tackles new tasks with the aid of manuals, databases, and other external tools rather than by relearning from scratch.}

\subsubsection{Pretraining}
In this stage SSL pretraining constructs pretext tasks from the data's structure and trains a large encoder on massive unlabeled datasets. Three categories of SSL strategies are most prevalent \cite{gui2024survey}.
\begin{itemize}
    \item \textbf{Generative SSL} trains the model to generate or reconstruct parts of the data from other parts, so that the encoder learns representations capturing the underlying statistical structure of the input. Typical instantiations include masked modeling, autoregressive modeling, and denoising or diffusion-based modeling \cite{gui2024survey}.
    \item \textbf{Contrastive SSL} operates in the latent space: semantically similar samples (positive pairs) are pulled closer together, while dissimilar samples (negative pairs) are pushed apart, encouraging the encoder to learn discriminative representations that preserve semantic similarity.
    \item \textbf{Hybrid SSL} combines generative and contrastive objectives to leverage their complementary strengths, yielding representations that are both context-aware and semantically discriminative.
\end{itemize}
Because these SSL objectives are applied to large and diverse unlabeled datasets, the resulting encoder captures rich, transferable representations that encode the statistical structure of the domain rather than overfitting to any single task.

\subsubsection{Adaptation}
Adaptation specializes the pretrained encoder for a downstream task using a small amount of labeled data. The encoder may be frozen with only a task head trained, or updated jointly with the head via full or parameter-efficient fine-tuning (PEFT).
\begin{itemize}
    \item \textbf{Full fine-tuning} trains both the pretrained encoder and a task-specific decoder end-to-end on a downstream labeled dataset. It offers the strongest adaptation but is computationally expensive and risks catastrophic forgetting.
    \item \textbf{PEFT} updates only a small fraction of parameters, reducing computational cost. Representative methods include partial fine-tuning that freezes early layers, low-rank adaptation (LoRA) injecting low-rank trainable matrices, and adapter layers inserting small bottleneck modules.
\end{itemize}
The choice among these strategies depends on the available labeled data, computational budget, and the degree of domain shift between pretraining and deployment.

\subsubsection{Utilization}
This stage resembles an expert tackling new tasks with external tools and knowledge. Utilization deploys the pretrained or adapted FM at inference time and can further enhance performance by injecting external context. The following techniques are widely used.
\begin{itemize}
    \item \textbf{Prompt engineering} augments the input with additional task-relevant signals that guide the frozen FM toward the desired behavior. In language models, such signals are typically hand-crafted text instructions or structured templates. More generally, prompt engineering injects auxiliary information into the FM through the input interface, either directly or via an auxiliary encoder.
    \item \textbf{In-context learning (ICL)} provides the FM with a few reference examples embedded directly in the input, from which the model infers the task pattern and produces the desired output for the new query. This allows the model to learn from a handful of demonstrations at inference time, without any parameter update.
    \item \textbf{Retrieval-augmented generation (RAG)} extends ICL by maintaining an external database of pre-stored reference examples. At inference time, a retrieval module fetches the most relevant ones as in-context references. Since task-specific knowledge resides in the database rather than the model, it can be updated without retraining.
\end{itemize}
Utilization techniques are particularly attractive for dynamic deployment environments, where new data can be incorporated directly through the external context, thereby enabling strong cross-scenario transfer, few-shot and zero-shot generalization, and seamless knowledge updating.

\begin{table*}[t]
\centering
\caption{Comparison of wireless localization paradigms.}
\label{tab:comparison}
\renewcommand{\arraystretch}{1.1}
    \begin{tabular}{|>{\raggedright\arraybackslash}p{2.3cm}|>{\raggedright\arraybackslash}p{2.7cm}|>{\raggedright\arraybackslash}p{3.3cm}|>{\raggedright\arraybackslash}p{4.0cm}|>{\raggedright\arraybackslash}p{3.6cm}|}
\hline
\textbf{Paradigm} & \textbf{Label Requirement} & \textbf{Generalization / Accuracy} & \textbf{Key Limitation} & \textbf{Relation to FM Paradigm} \\
\hline
Model-Based & None (model-driven) & High in LoS / calibrated wideband or massive-MIMO; degrades under severe NLoS & Requires accurate model and tight synchronization; suffers from limited resolution, multi-bounce, and LoS blockage. & FM features can complement geometric estimators (e.g., ToA/AoA-based) \\
\hline
Supervised Learning & Large labeled dataset per environment & High in-domain; poor across environments & Catastrophic failure in new environments & FM provides transferable pretrained features that replace per-environment retraining
 \\
\hline
Channel Charting & None for charting; labels for anchoring & Unsupervised representation learning; extends toward universal CSI embeddings & Lacks absolute coordinate anchoring and robust cross-scenario calibration & FM extends CC-style embeddings with pretraining; adds absolute anchoring \\
\hline
Channel Knowledge Map & Location-indexed channel knowledge (site-specific prior) & Good within the mapped area & Construction and update cost as the environment changes & 
Site-specific prior injected into the FM as retrieval context or prompts \\
\hline
\textbf{Foundation Model} & \textbf{Few labeled samples} & \textbf{Strong cross-environment; high accuracy} & \textbf{Pretraining compute cost} & \textbf{---} \\
\hline
\end{tabular}
\end{table*}

\subsection{Paradigm Comparison and Key Advantages}
\label{subsec:comparison}

Wireless localization has been approached from several distinct paradigms. Before presenting the FM-based approach, we briefly review the existing paradigms and highlight their key limitations, as summarized in Table~\ref{tab:comparison}:

\subsubsection{Model-based methods} estimate geometric channel parameters such as ToA and AoA using parametric signal processing, based on high-resolution methods such as multiple signal classification (MUSIC). They require no training data, but their accuracy is limited by the available bandwidth and aperture and by stringent synchronization requirements, and it degrades in complex NLoS scenarios with dense multi-bounce propagation or LoS blockage, where the parametric model no longer matches the true channel.

\subsubsection{Supervised learning (SL)} approaches, such as fingerprinting, learn direct CSI-to-position mappings from large labeled datasets. They achieve high in-domain accuracy but memorize environment-specific fingerprint patterns, leading to catastrophic degradation in new environments.

\subsubsection{Channel charting (CC)} {is an unsupervised representation-learning paradigm that maps unlabeled CSI into a low-dimensional embedding space preserving spatial neighborhood structure, rather than directly optimizing for localization. Classical CC produces relative topology rather than absolute positions, requiring post-hoc anchoring, while recent work extends CC toward more universal CSI embeddings. Its main limitations lie in absolute coordinate anchoring and cross-scenario calibration.}

\subsubsection{Channel knowledge maps (CKMs)} {provide site-specific, location-indexed channel knowledge (e.g., path loss, delay spread, or full CSI) that serves as an environment prior. Such priors can be constructed through ray-tracing, dense measurements, digital twins, interpolation, or generative and parametric environment-aware models. Although originally designed to predict channel knowledge from location for environment-aware communication, CKMs can also be exploited for localization, e.g., by matching a new measurement against the prior to retrieve the most likely position. Their main cost lies in constructing and updating the prior as the environment changes or coverage extends.}

{By combining the strengths of the above paradigms, the FM paradigm offers a unifying framework that inherits the label-free representation learning of channel charting, the context injection of CKMs, and the direct CSI-to-position mapping of supervised fingerprinting within a single pretraining–adaptation–utilization pipeline. Additionally, although FM-based localization is less interpretable than model-based estimators, its pretraining objective drives the FM to learn the underlying structure and propagation knowledge of the channel, which to some extent compensates for this lack of explicit modeling. Building on these properties, FM-based localization provides the following key advantages.}

\begin{itemize}
\item \textbf{Cross-environment generalization.} By learning general-purpose representations from diverse channel datasets, FMs capture invariant propagation characteristics that persist across environments and system configurations. Therefore, a single pretrained FM encoder can be reused across multiple deployment scenarios, significantly improving scalability \cite{yang2025generative}.

\item \textbf{Label-efficient adaptation.} Leveraging the rich knowledge embedded during pretraining, FM-based localization can adapt to new environments in few-shot or even zero-shot settings. This substantially reduces the data collection effort and deployment overhead compared to SL and CKM approaches \cite{pan2025large}.

\item \textbf{Shared representations across channel-related tasks.} Since localization, communication, and sensing share the same wireless propagation mechanisms, a single pretrained FM can yield channel representations reusable across these tasks.\footnote{We primarily focus on wireless localization in this article.} Accurate location estimates can in turn enhance communication functions such as beam management and channel prediction, creating a bidirectional synergy \cite{guo2026large}.

\item \textbf{Multi-BS and multi-modal fusion.} FMs fuse observations from multiple BSs and heterogeneous modalities (e.g., radio, vision, inertial) within a unified representation space, where cross-modal alignment during pretraining enables seamless integration of complementary information. When some modalities become unavailable, the FM can still perform localization using the remaining observations, enhancing system resilience \cite{si2025cross}.
\end{itemize}

\section{FM-based Framework For Wireless Localization}\label{sec:framework}

\begin{figure*}[t]
    \centering
    \includegraphics[scale=0.52]{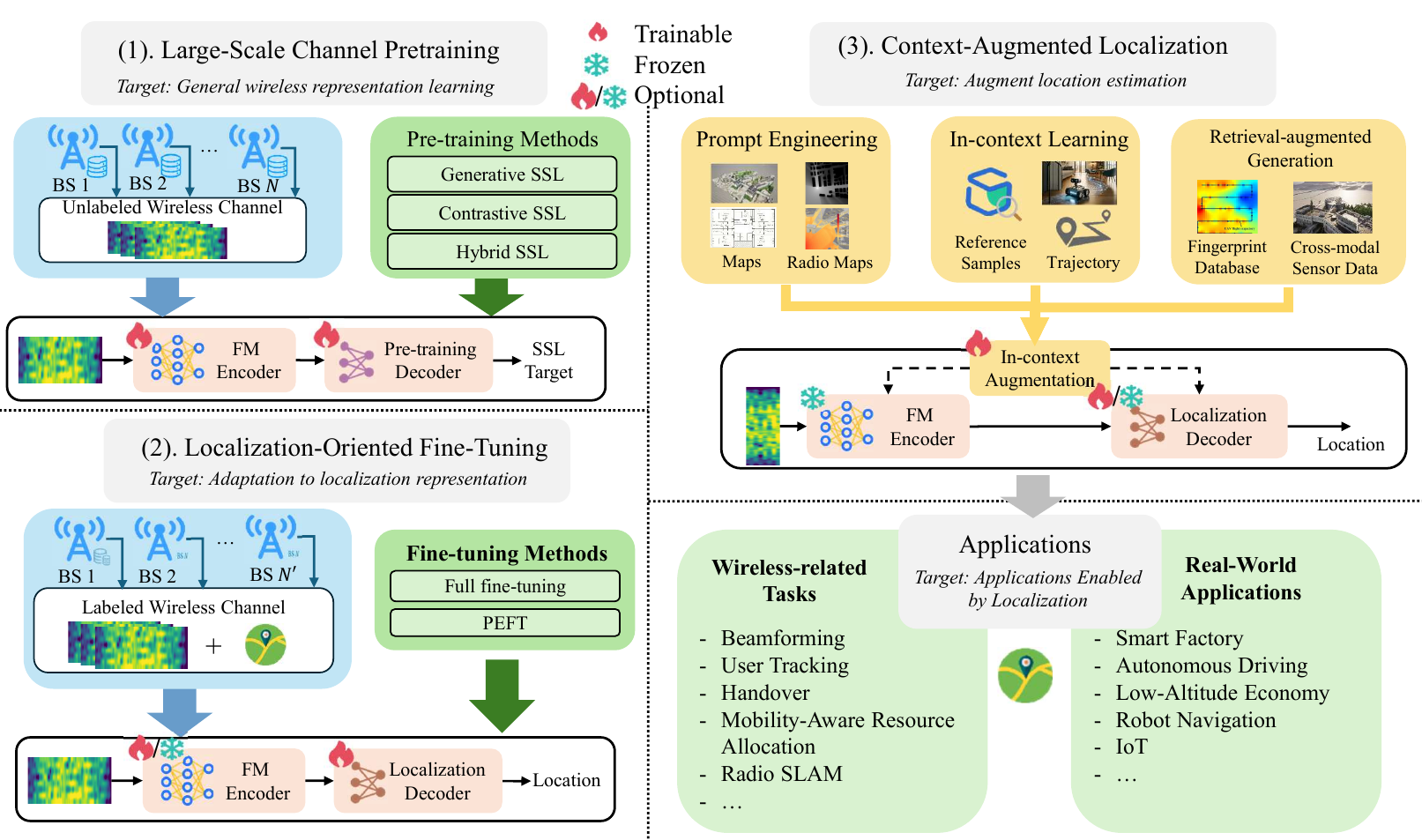}
    \caption{{The proposed three-stage FM-based localization framework: (1) large-scale channel pretraining, (2) localization-oriented fine-tuning, and (3) context-augmented inference, together with the location-aware applications they enable.}}
    \label{fig:framework}
    \vspace{-3mm}
\end{figure*}

As illustrated in Fig.~\ref{fig:framework}, the proposed FM-based localization framework consists of 3 progressive stages together with the location-aware applications they enable.

\subsubsection{Stage 1: Large-Scale Channel Pretraining}

The FM-based localization system begins with SSL pretraining on large-scale unlabeled CSI data collected from diverse environments, BS deployments, and system configurations (e.g., different frequency bands, antenna geometries, and bandwidths). The objective is not to perform localization directly, but to learn general-purpose channel representations that capture invariant propagation structures shared across different settings, yielding effective features for wireless channels. These representations encode fundamental channel properties and benefit not only localization but also other channel-related tasks such as channel estimation, beam prediction, and sensing. Since no labels are required, acquiring such datasets is technically straightforward. Meanwhile, the diversity of the collected data has a direct impact on the quality of the extracted features. Therefore, the pretraining dataset should cover as many heterogeneous deployment scenarios as possible, spanning a wide range of environments, BS configurations, and propagation conditions, so that the learned representations can generalize well to unseen settings. In addition, the effectiveness of the learned representations also depends on the choice of SSL algorithm, as different SSL objectives yield representations with varying strengths for different downstream tasks. A detailed survey of SSL algorithms for wireless channels can be found in~\cite{jiang2025towards}.

\subsubsection{Stage 2: Localization-Oriented Fine-Tuning}

Stage 2 consists of the pretrained FM encoder and a localization-oriented decoder. {Since the SSL representations are not specifically tailored for localization, the encoder can either be kept frozen, with only the decoder trained on labeled CSI-position datasets, or be fine-tuned jointly with it. Freezing is the most lightweight option and fully preserves the pretrained representations, but may leave a noticeable gap compared to a fully task-adapted model. Joint fine-tuning closes this gap by shifting the representations from task-general to localization-oriented at the cost of additional training. The choice depends on the available labeled data, the computational budget, and whether the backbone must remain shared with other downstream tasks. Such labeled data can be collected from existing deployed networks, and thanks to the strong representation inherited from Stage 1, only a small amount is required. When fine-tuning is adopted, the strategy can be flexibly chosen.} Full fine-tuning updates all encoder parameters for maximum adaptation. Alternatively, PEFT methods such as LoRA or Adapter layers can be applied to preserve pretrained knowledge while reducing computational cost. During inference, the incoming CSI is fed into the FM encoder to obtain channel representations, which, together with BS configuration information, are passed through the localization decoder to produce position estimates. However, at this stage, the localization-oriented representations remain scenario-agnostic. They do not account for the specific propagation characteristics of a particular deployment environment, such as local scatterer distributions, building geometry, or site-specific NLoS conditions. As a result, while the model already delivers reasonable localization performance, there is still room for improvement by incorporating environment-specific information.

\subsubsection{Stage 3: Context-Augmented Localization}

After Stage 2 produces the localization-oriented representations, Stage 3 can either directly perform inference using the FM and the decoder from Stage 2, or exploit the opportunity to jointly leverage auxiliary information for further accuracy improvement. Such auxiliary information includes site-specific knowledge (e.g., floor plans, radio maps, deployment metadata), heterogeneous sensing observations (e.g., cameras, LiDAR, inertial sensors), local CSI-position databases, and trajectory history. {In the former case, the decoder is reused as is with no additional training; in the latter, it is extended or replaced by a context-aware decoder that jointly processes the channel representations and the injected context, while the FM encoder can remain frozen. The utilization techniques, such as prompt engineering, ICL, and RAG, provide effective mechanisms to integrate these auxiliary signals into the FM-based localization:}
\begin{itemize}
    \item \textbf{Prompt engineering:} Site-specific knowledge includes floor plans, radio maps from digital twins, and deployment metadata. Such knowledge can be encoded via lightweight auxiliary encoders and injected into the FM as prompt-like representations, which condition the FM on local propagation characteristics. Since the prompts are static, they can be precomputed offline and reused across all queries in the same scenario. Different scenarios are then supported by swapping the corresponding prompts.
    \item \textbf{ICL:} A small number of reference samples and recent trajectory segments observed in the current scenario can be provided directly alongside the input as demonstrations, guiding the FM to produce positioning outputs consistent with the local context. In this way, when migrating to a new scenario or when the environment changes, it suffices to update the reference set, without any model retraining.
    \item \textbf{RAG:} External sources are queried at inference time to fetch relevant references. Same-modality retrieval, e.g., nearest-neighbor CSI-position pairs from a local database of the target environment, provides scenario-specific context to the FM and integrates naturally with ICL by supplying the retrieved samples as demonstrations. Cross-modal retrieval can further enable multi-modal fusion by incorporating real-time visual or LiDAR observations from co-located sensors. {Unlike fingerprinting and CKM lookup, where retrieval itself acts as the estimator, FM-RAG retrieves in a pretrained representation space and uses the retrieved samples as context to an adapted FM. The local database thus acts as an updatable environment memory, while the FM provides transferable representations across scenarios. Note that a labeled local database incurs a collection cost comparable to CKM construction. Its advantage lies in representation reuse and update efficiency rather than in reduced target-environment data.}
\end{itemize}
{Unlike LLMs where utilization operates on a fully frozen FM, the wireless setting often requires the FM to first learn how to consume such auxiliary signals, e.g., by fine-tuning a small subset of parameters or training lightweight modules (auxiliary encoders for site-specific priors, fusion heads for retrieved references). Such adaptation is nonetheless far lighter than Stage 1 pretraining, and once completed, deployment in new scenarios proceeds without further weight updates.}

\subsubsection{Location-Aware Applications}

{The position estimates and location-aware representations obtained from the previous stages can be leveraged to optimize a wide range of applications, where the reusable representations further allow channel-related tasks to share a single FM backbone.} Within wireless systems, accurate location information enables predictive beamforming by anticipating user movement, facilitates user tracking for sustained connectivity, supports proactive handover based on trajectory prediction, and allows mobility-aware resource allocation that reduces interference and improves spectral efficiency. Location information together with channel representations also fits naturally into radio simultaneous localization and mapping (SLAM) \cite{lotti2023radio}, where the system jointly estimates user positions and builds a map of the radio environment.

Beyond wireless system optimization, FM-based localization serves as a key enabler for emerging location-critical applications, such as robot coordination in smart factories, positioning in autonomous driving under GNSS-denied conditions, reliable localization for unmanned aerial vehicles in the low-altitude economy, indoor robot navigation, and scalable localization for large-scale IoT deployments. Moreover, the location-aware features extracted by the FM can be seamlessly integrated with other application-level FMs, such as vision-language models, world models, or embodied AI agents, enabling end-to-end cross-domain intelligence in which wireless sensing and high-level reasoning jointly drive the task. In this way, FM-based localization further paves the way toward AI-native wireless networks, where the localization capabilities of the wireless infrastructure and the needs of upper-layer applications are tightly integrated through a shared FM.

\section{Case study}

To {provide an illustrative case study}, we conduct two experiments that respectively demonstrate the key advantages of FM-based localization: (i) the benefit of diverse channel pretraining for label-efficient adaptation, and (ii) the generalizability of the proposed framework. We generate multi-scenario CSI-position datasets using Sionna, a ray-tracing-based wireless simulation platform. Each scenario corresponds to a distinct propagation environment, differing in BS orientation, bandwidth configuration, and building topology, and covers a $32\,\mathrm{m}\times 32\,\mathrm{m}$ area. For each scenario, 10\,000 CSI samples are randomly collected at user positions for unlabeled pretraining, while labeled CSI-position pairs are drawn separately for fine-tuning and testing. The FM encoder adopts a lightweight transformer backbone pretrained via masked channel reconstruction, and a lightweight multi-layer perceptron (MLP) is attached as the localization decoder.

\begin{figure}[t]
    \centering
    \begin{tikzpicture}
        \begin{axis}[
            width=0.9\columnwidth,
            height=0.6\columnwidth,
            xlabel={Number of labeled fine-tuning samples},
            ylabel={Mean localization error (m)},
            xmin=500, xmax=2000,
            ymin=1.0, ymax=5,
            xtick={500,1000,1500,2000},
            ytick={1,2,3,4,5},
            grid=both,
            major grid style={line width=0.2pt,draw=gray!30},
            minor grid style={line width=0.1pt,draw=gray!15},
            tick label style={font=\footnotesize},
            label style={font=\footnotesize},
            legend style={font=\footnotesize, at={(0.99,0.99)}, anchor=north east, draw=gray!50},
        ]
            \addplot[color=black, mark=o, mark size=2pt, line width=1.0pt] coordinates {
                (500, 4.71)
                (1000, 3.81)
                (1500, 3.28)
                (2000, 2.93)
            };
            \addlegendentry{SL}        
            \addplot[color=red!70!black, mark=square*, mark size=2pt, line width=1.0pt] coordinates {
                (500, 3.10)
                (1000, 2.54)
                (1500, 2.22)
                (2000, 2.03)
            };
            \addlegendentry{$K=1$ scenario}
            \addplot[color=blue!70!black, mark=*, mark size=2pt, line width=1.0pt] coordinates {
                (500, 1.86)
                (1000, 1.58)
                (1500, 1.40)
                (2000, 1.35)
            };
            \addlegendentry{$K=10$ scenarios}
            \addplot[color=green!50!black, mark=triangle*, mark size=2.5pt, line width=1.0pt] coordinates {
                (500, 1.80)
                (1000, 1.54)
                (1500, 1.32)
                (2000, 1.22)
            };
            \addlegendentry{$K=100$ scenarios}
        \end{axis}
    \end{tikzpicture}
    \caption{Impact of pretraining scenario diversity on label-efficient localization. As the number of pretraining scenarios $K$ increases, the mean localization error after fine-tuning drops consistently across all labeled-data budgets.}
    \label{fig:exp1}
\end{figure}
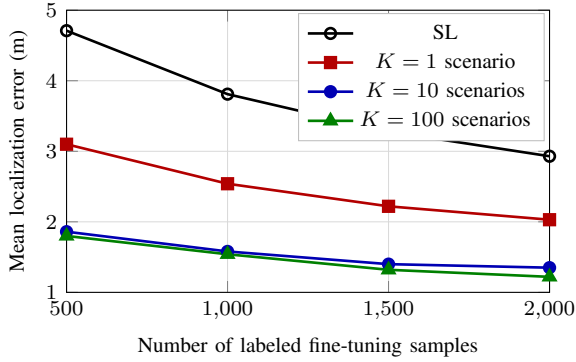

\subsection{Effect of Pretraining Scenario Diversity}

This experiment examines how the diversity of pretraining data affects downstream label-efficient localization. We pretrain the FM encoder on $K\in\{1,10,100\}$ distinct scenarios. The pretrained encoder is then fine-tuned on one additional scenario using 500, 1000, 1500, or 2000 labeled CSI-position pairs, and evaluated on 1000 test samples from that scenario. As a baseline, we also train the same network directly on the labeled data without pretraining.
Fig.~\ref{fig:exp1} reports the mean localization error with different numbers of labeled fine-tuning samples. Fig.~\ref{fig:exp1} shows that pretraining consistently outperforms training from scratch. With 500 labels, the model pretrained on $K=100$ scenarios attains a 1.80\,m error, which corresponds to a 62\% reduction from the no-pretraining baseline at 4.71\,m and a 42\% reduction from the $K=1$ pretrained model at 3.10\,m. Notably, however, going from $K=10$ to $K=100$ only marginally further reduces the error, by about 3\% under 500-label fine-tuning, suggesting that the gain from further increasing scenario diversity tends to saturate. This saturation may partly reflect the limited channel diversity of our Sionna-based pretraining datasets, and constructing richer channel datasets that combine ray-tracing simulations with real-world measurements across diverse frequency bands and deployment scenarios remains an important direction for further unlocking the benefits of FM pretraining.

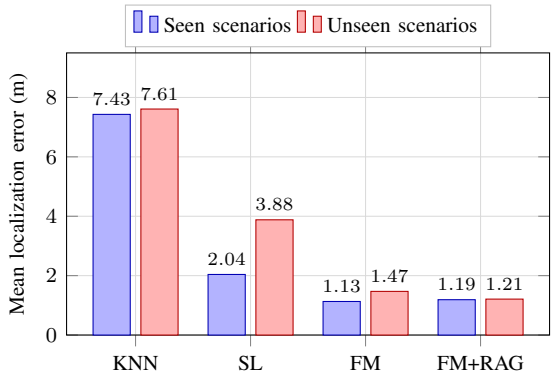
\begin{figure}[t]
    \centering
    \begin{tikzpicture}
        \begin{axis}[
            width=0.9\columnwidth,
            height=0.6\columnwidth,
            ybar=4pt,
            bar width=14pt,
            enlarge x limits=0.2,
            ymin=0,
            ymax=9.5,
            ylabel={Mean localization error (m)},
            symbolic x coords={KNN, SL, FM, FM+RAG},
            xtick=data,
            ytick={0,2,4,6,8},
            tick label style={font=\footnotesize},
            label style={font=\footnotesize},
            legend style={font=\footnotesize, at={(0.5,1.02)}, anchor=south, legend columns=2, draw=gray!50},
            grid=major,
            major grid style={line width=0.2pt,draw=gray!30},
            nodes near coords,
            nodes near coords style={font=\scriptsize, /pgf/number format/.cd, fixed, precision=2, /tikz/.cd},
        ]
            \addplot[fill=blue!30, draw=blue!70!black] coordinates {
                (KNN, 7.43)
                (SL, 2.04)
                (FM, 1.13)
                (FM+RAG, 1.19)
            };
            \addplot[fill=red!30, draw=red!70!black] coordinates {
                (KNN, 7.61)
                (SL, 3.88)
                (FM, 1.47)
                (FM+RAG, 1.21)
            };
            \legend{Seen scenarios, Unseen scenarios}
        \end{axis}
    \end{tikzpicture}
    \caption{Cross-environment localization performance under 4 schemes.}
    \label{fig:exp2}
\end{figure}

\subsection{Cross-Environment Generalization}

This experiment evaluates the FM-based framework in seen scenarios included in fine-tuning versus unseen scenarios that are not. Starting from the FM pretrained on 100 scenarios, we fine-tune it on 10 randomly selected scenarios, each providing 2000 labeled CSI-position samples. For seen scenarios, we draw 1000 additional test samples from the fine-tuned scenarios; for unseen scenarios, we generate 10 entirely new scenarios. Four schemes are compared: (1) KNN, which directly matches the query CSI against a local database via channel similarity and outputs the position of the nearest samples; (2) SL, a supervised fingerprinting baseline trained directly on the labeled data without pretraining; (3) FM, with localization-oriented fine-tuning; and (4) FM+RAG, which further retrieves from a local database of 2000 labeled samples via KNN in the FM representation space.
As shown in Fig.~\ref{fig:exp2}, on seen scenarios both FM and FM+RAG substantially outperform the KNN and SL baselines. On unseen scenarios, SL degrades sharply to 3.88 m, and the FM error rises to 1.47 m owing to the scenario-agnostic nature of its representations. KNN, in contrast, performs almost identically in both cases, since it involves no training and its accuracy is governed by the database density rather than by scenario familiarity. FM+RAG recovers most of the lost accuracy, achieving 1.21 m. As it differs from KNN only in the space where similarity is measured, the gap between them shows that the FM provides a stronger representation, enabling far more informative neighbors to be retrieved from the same database. These results show that retrieval-based context augmentation is valuable in unseen environments, where it closes the gap without retraining the backbone.

\section{OUTLOOK}
\label{sec:opportunities}

The FM paradigm opens several research directions for wireless localization, each carrying both clear potential and open challenges.

\subsubsection{FM representations across modalities and configurations} Beyond CSI, 6G networks will co-exist with diverse sensing modalities, including camera images, LiDAR point clouds, radar, and inertial measurements. Jointly pretraining the FM on multiple modalities to learn aligned cross-modal representations would enable seamless fusion at inference time, where visual or geometric context helps resolve ambiguities in NLoS CSI. Even within the radio modality, however, BSs operate under diverse configurations (antenna geometry, carrier frequency, bandwidth, subcarrier spacing), and representations must be robust to such differences to be truly reusable across deployments. Achieving this requires careful pretraining strategies, such as configuration augmentation or contrastive objectives that explicitly encourage invariance.

\subsubsection{Uncertainty-aware and trustworthy localization} FM-based localization currently produces point estimates, but safety-critical applications such as autonomous driving and UAV navigation require calibrated uncertainty. Extending the FM output to a probabilistic distribution through generative modeling (e.g., conditional flow matching or diffusion-based decoders) would enable risk-aware decisions, such as triggering fallbacks when the model is not confident \cite{liu2025difflow3d}. Calibrated uncertainty is also a prerequisite for trustworthiness more broadly. Reliable behavior under adversarial perturbations, hardware impairments, and out-of-distribution inputs further calls for robust pretraining and run-time anomaly detection.

\subsubsection{Efficient deployment across edge, BS, and cloud} Transformer-based encoders have non-trivial inference costs that may exceed the latency budget of edge devices. Model compression techniques (such as knowledge distillation, quantization, pruning, and efficient attention variants), together with split inference across edge, BS, and cloud, are essential to meet stringent latency requirements without significantly degrading representation quality. Such a split also opens the possibility of sharing a single backbone across BSs while keeping only lightweight task heads at the edge, turning a deployment constraint into an architectural design choice.

\subsubsection{Organization and integration of external context} Utilization techniques such as prompt engineering, ICL, and RAG inject external knowledge into the FM at inference time, but their effectiveness hinges on how that knowledge is organized and represented. Site-specific priors (e.g., floor plans, radio maps, deployment metadata) must be encoded into compact, FM-compatible representations through carefully designed auxiliary encoders. Reference databases for ICL and RAG should further balance coverage, freshness, and scale, raising open questions on hierarchical indexing, cross-BS organization, and cost-effective labeling.

\subsubsection{Agentic localization} A further direction is to develop FM-based agents for localization. Such an agent autonomously decides which tool to use, when to retrieve information, and whether further measurements are needed. When the current observation is insufficient, it could query a CKM or local fingerprint database, request additional BS or multi-modal measurements, and pass the resulting estimate and its uncertainty to downstream decisions. Here the FM supplies representation and reasoning capability, external databases supply knowledge, and the agent acts as the orchestration layer. Open problems include coordinating these components, defining the action space and termination criterion, and bounding the latency and signaling overhead of extra measurement requests.

\subsubsection{Privacy-preserving and continually updated learning} Pretraining benefits from data diversity that often spans multiple operators, deployments, and user populations, which typically raises privacy concerns~\cite{wang2025federated}. Federated pretraining can enable multi-operator collaboration without sharing raw CSI, and differential privacy can further protect user location information during both pretraining and local database construction for RAG. Keeping the model current poses a related problem, as real-world environments evolve due to construction, furniture rearrangement, or seasonal propagation changes. FM-based localization can adapt through two complementary mechanisms~\cite{yu2026recent}, namely incrementally updating the RAG database with newly collected measurements and periodically performing lightweight continual pretraining on fresh unlabeled CSI, though doing so under privacy constraints and without catastrophic forgetting remains open.

\subsubsection{Standardization, benchmarking, and 3GPP integration} The 3GPP NR positioning architecture has standardized AI-based positioning enhancements, treating localization as a representative use case of AI in wireless networks~\cite{pan2025ai}. The proposed framework can further serve as a shared backbone for other 3GPP-defined AI tasks such as CSI feedback, beam management, and channel prediction, enabling joint optimization across multiple AI functions within a unified model. Realizing this at scale requires infrastructure the wireless community currently lacks, including standardized benchmarks, evaluation protocols, and shared pretrained model repositories. Establishing common datasets and metrics, along with open pretrained checkpoints, is essential to accelerate progress and ensure fair comparison.

\color{black}

\section{Conclusion}

In this article, we reviewed the fundamentals of FMs, compared existing wireless localization paradigms with the FM-based paradigm and introduced a unified FM-based localization framework. Through case studies, we observed that increasing the diversity of pretraining scenarios consistently reduces the labeled data required to reach a given accuracy, and that retrieval-based context augmentation can recover most of the accuracy loss when the model is deployed in previously unseen scenarios. Finally, we presented an outlook on the key research directions for advancing FM-based wireless localization toward large-scale deployment.

\bibliographystyle{IEEEtran}
\bibliography{IEEEabrv, ref}

\begin{thebibliography}{10}
\providecommand{\url}[1]{#1}
\csname url@samestyle\endcsname
\providecommand{\newblock}{\relax}
\providecommand{\bibinfo}[2]{#2}
\providecommand{\BIBentrySTDinterwordspacing}{\spaceskip=0pt\relax}
\providecommand{\BIBentryALTinterwordstretchfactor}{4}
\providecommand{\BIBentryALTinterwordspacing}{\spaceskip=\fontdimen2\font plus
\BIBentryALTinterwordstretchfactor\fontdimen3\font minus \fontdimen4\font\relax}
\providecommand{\BIBforeignlanguage}[2]{{%
\expandafter\ifx\csname l@#1\endcsname\relax
\typeout{** WARNING: IEEEtran.bst: No hyphenation pattern has been}%
\typeout{** loaded for the language `#1'. Using the pattern for}%
\typeout{** the default language instead.}%
\else
\language=\csname l@#1\endcsname
\fi
#2}}
\providecommand{\BIBdecl}{\relax}
\BIBdecl

\bibitem{chen2022tutorial}
H.~Chen, H.~Sarieddeen \emph{et~al.}, ``A tutorial on terahertz-band localization for {6G} communication systems,'' \emph{IEEE Commun. Surveys Tuts.}, vol.~24, no.~3, pp. 1780--1815, 2022.

\bibitem{pan2025ai}
G.~Pan, Y.~Gao \emph{et~al.}, ``{AI}-driven wireless positioning: Fundamentals, standards, state-of-the-art, and challenges,'' \emph{IEEE Commun. Surveys Tuts.}, vol.~28, pp. 4394--4428, 2026.

\bibitem{fei2022towards}
N.~Fei, Z.~Lu \emph{et~al.}, ``Towards artificial general intelligence via a multimodal foundation model,'' \emph{Nature Commun.}, 2022.

\bibitem{guo2026large}
J.~Guo, Y.~Cui \emph{et~al.}, ``Large {AI} models for wireless physical layer,'' \emph{{IEEE} Commun. Mag.}, vol.~64, no.~5, pp. 148--155, 2026.

\bibitem{jiang2025towards}
J.~Jiang, Y.~Gao \emph{et~al.}, ``Towards channel foundation models ({CFMs}): Motivations, methodologies and opportunities,'' \emph{arXiv preprint arXiv:2507.13637}, 2025.

\bibitem{pan2025large}
G.~Pan, K.~Huang \emph{et~al.}, ``Large wireless localization model ({LWLM}): A foundation model for positioning in {6G} networks,'' \emph{arXiv preprint arXiv:2505.10134}, 2025.

\bibitem{liu2026self}
Y.~Liu, H.~Si \emph{et~al.}, ``A self-supervised masked autoencoder leveraging temporal-frequency representation for {CSI} localization,'' \emph{IEEE Trans. Netw. Sci. Eng.}, 2026.

\bibitem{si2025cross}
H.~Si, X.~Guo \emph{et~al.}, ``Cross-scenario foundation localization models: Architecture, key technologies, and challenges,'' \emph{{IEEE} Wireless Commun.}, 2025.

\bibitem{zhao2026survey}
W.~X. Zhao, K.~Zhou \emph{et~al.}, ``A survey of large language models,'' \emph{Front. Comput. Sci.}, vol.~20, no.~12, p. 2012627, 2026.

\bibitem{gui2024survey}
J.~Gui, T.~Chen \emph{et~al.}, ``A survey on self-supervised learning: Algorithms, applications, and future trends,'' \emph{{IEEE} Trans. Pattern Anal. Mach. Intell.}, vol.~46, no.~12, pp. 9052--9071, 2024.

\bibitem{yang2025generative}
Z.~Yang, G.~Chi \emph{et~al.}, ``Generative {AI} meets wireless sensing: Towards wireless foundation model,'' \emph{arXiv preprint arXiv:2509.15258}, 2025.

\bibitem{lotti2023radio}
M.~Lotti, G.~Pasolini \emph{et~al.}, ``Radio {SLAM} for {6G} systems at {THz} frequencies: Design and experimental validation,'' \emph{{IEEE} J. Sel. Topics Signal Process.}, vol.~17, no.~4, pp. 834--849, 2023.

\bibitem{liu2025difflow3d}
J.~Liu, W.~Ye \emph{et~al.}, ``{DifFlow3D}: Toward robust uncertainty-aware scene flow estimation with diffusion model,'' \emph{{IEEE} Trans. Pattern Anal. Mach. Intell.}, vol.~48, no.~3, pp. 2837--2854, 2025.

\bibitem{wang2025federated}
Z.~Wang, Y.~Zhou \emph{et~al.}, ``Federated fine-tuning for pre-trained foundation models over wireless networks,'' \emph{{IEEE} Trans. Wireless Commun.}, vol.~24, no.~4, pp. 3450--3464, 2025.

\bibitem{yu2026recent}
D.~Yu, X.~Zhang \emph{et~al.}, ``Recent advances of multimodal continual learning: A comprehensive survey,'' \emph{IEEE Trans. Neural Networks Learn. Syst.}, 2026.

\end{thebibliography}

\end{document}